\documentclass[11pt,twoside]{article}
\usepackage{asp2010}

\resetcounters

\begin{document}

\title{The extended atmospheres of Mira variables probed by VLTI, VLBA,
and APEX}
\author{M. Wittkowski$^1$, D. A. Boboltz$^2$, C. de Breuck$^1$, M. Gray$^3$,
E. Humphreys$^1$, M. Ireland$^4$, I. Karovicova$^1$, K. Ohnaka$^5$,
A. E. Ruiz-Velasco$^{1,6}$, M. Scholz$^{7,4}$, P. Whitelock$^{8,9}$,
and A. Zijlstra$^3$
\affil{$^1$ESO, Karl-Schwarzschild-Str. 2, 85748 Garching bei M\"unchen, 
Germany}
\affil{$^2$ US Naval Observatory, 3450 Massachusetts Avenue, NW, Washington, DC 20392-5420, USA}
\affil{$^3$ Jodrell Bank Centre for Astrophysics, School of Physics 
and Astronomy, University of Manchester, Manchester M13 9PL, UK}
\affil{$^4$ Sydney Institute for Astronomy, School of Physics, University
of Sydney, Sydney, NSW 2006, Australia}
\affil{$^5$ Max-Planck-Institut f\"ur Radioastronomie, Auf dem H\"ugel 69, 53121 Bonn, Germany}
\affil{$^6$ Departamento de Astronom{\'i}a, Universidad de Guanajuato, Apartado Postal 144, 
36000 Guanajuato, Mexico}
\affil{$^7$ Zentrum f\"ur Astronomie (ZAH), Institut f\"ur Theoretische 
Astrophysik, Albert Ueberle-Str. 2, 69120 Heidelberg, Germany}
\affil{$^8$ South African Astronomical Observatory, PO Box 9, 7935 Observatory,
South Africa}
\affil{$^9$ University of Cape Town, 7701 Rondebosch, South Africa}
}

\begin{abstract}
We present an overview on our project to study the extended atmospheres and
dust formation zones of Mira stars using coordinated observations with 
the Very Large Telescope Interferometer (VLTI), the Very Long Baseline
Array (VLBA), and the Atacama Pathfinder Experiment (APEX). 
The data are interpreted using an approach of combining recent
dynamic model atmospheres with a radiative transfer model of the dust shell,
and combining the resulting model structure with a maser propagation model.
\end{abstract}

\section{Introduction and project outline}
Mass-loss becomes
increasingly important toward the tip of the AGB evolution.
While the mass-loss process during the AGB phase is the most important
driver for the further stellar evolution toward the PN phase, the details
of the mass-loss process and its connection to the structure of the
extended atmospheres and the stellar pulsation are not well understood and are
currently a matter of debate.

Here, we present an overview on our established project of coordinated
interferometric observations at infrared and radio wavelengths.
Our goal is to establish
the radial structure and kinematics of the stellar atmosphere and the
circumstellar environment to understand better the mass-loss process
and its connection to stellar pulsation.
We also aim at tracing asymmetric structures from small to large
distances in order to constrain shaping processes during the
AGB evolution, which may lead to the observed diversity of shapes of
planetary nebulae.
We use two of the highest resolution interferometers in the world,
the Very Large Telescope Interferometer (VLTI) and the Very Long Baseline
Array (VLBA) to study AGB stars and their circumstellar envelopes
from near-infrared to radio wavelengths. For some sources, we have
included near-infrared broad-band photometry obtained at the
South African Astronomical Observatory (SAAO) in order to derive
effective temperature values. 
We have started to
use the Atacama Pathfinder Experiment (APEX) to investigate the line strengths
and variability of high frequency SiO maser emission,

\section{Observations}
\label{sec:observations}
Our pilot study included coordinated observations of the Mira variable
S Ori including VINCI $K$-band measurements at the VLTI and SiO maser
measurements at the VLBA (Boboltz \& Wittkowski 2005). For the Mira
variables S Ori, GX Mon, RR Aql and the supergiant AH Sco, we obtained
long-term mid-infrared interferometry covering several pulsation cycles
using the MIDI instrument at the VLTI
coordinated with VLBA SiO (42.9 GHz and 43.1 GHz transitions) observations 
(Wittkowski et al. 2007; Karovicova et al, these proceedings).
For the Mira variables R Cnc and X Hya, we
coordinated near-infrared interferometry (VLTI/AMBER), 
mid-infrared interferometry (VLTI/MIDI), VLBA/SiO maser observations,
VLBA/H$_2$O maser, and near-infrared photometry at the SAAO
(work in progress).
Most recently, measurements of the $v=1$ and $v=2$ $J=7-6$ SiO maser
transitions toward our program stars were obtained at two epochs
using APEX (work in progress).
\section{Modeling}
\label{sec:modeling}
The P and M model series by Ireland et al. (2004a/b)
were chosen as the currently best available
option to describe dust-free Mira star atmospheres.
Wittkowski et al.~(2007)
have added an ad-hoc radiative transfer model to these model series 
to describe
the dust shell as observed with the mid-infrared interferometric 
instrument MIDI using the radiative transfer code {\tt mcsim\_mpi}
by Ohnaka et al. (2007).
Gray et al. (2009) have combined
these hydrodynamic atmosphere plus dust shell models with a maser 
propagation code in order
to describe the SiO maser observations. 
Most recently, we have also used new dynamic atmosphere series (CODEX series)
by Ireland et al. (2008), which use the opacity sampling method, and are
available for additional stellar parameters compared to the P/M series.
\section{Results}
The pilot study on the Mira variable S Ori (Boboltz \& Wittkowski 2005)
revealed that the SiO maser ring radii lie at 2.0 R$_\mathrm{cont}$ (43.1 GHz 
transition) and 1.9 R$_\mathrm{cont}$ (42.8 GHz transition) at phase 0.7.
The stellar diameter was estimated by measuring the uniform disk diameter
and correcting it for the continuum diameter using dynamic model
atmospheres as described in Sect.~\ref{sec:modeling}. This result is free
of the usual uncertainty inherent in comparing observations widely
spaced in phase and/or using directly uniform disk diameters which may
be contaminated by extended molecular layers.
Fedele et al. (2005) estimated in an analogous way SiO maser ring
radii of 1.9 R$_\mathrm{cont}$ and 1.8 R$_\mathrm{cont}$, respectively,
for the Mira variable R Leo at phase 0.1.  

\begin{figure}
\centering
\resizebox{0.45\hsize}{!}{\includegraphics{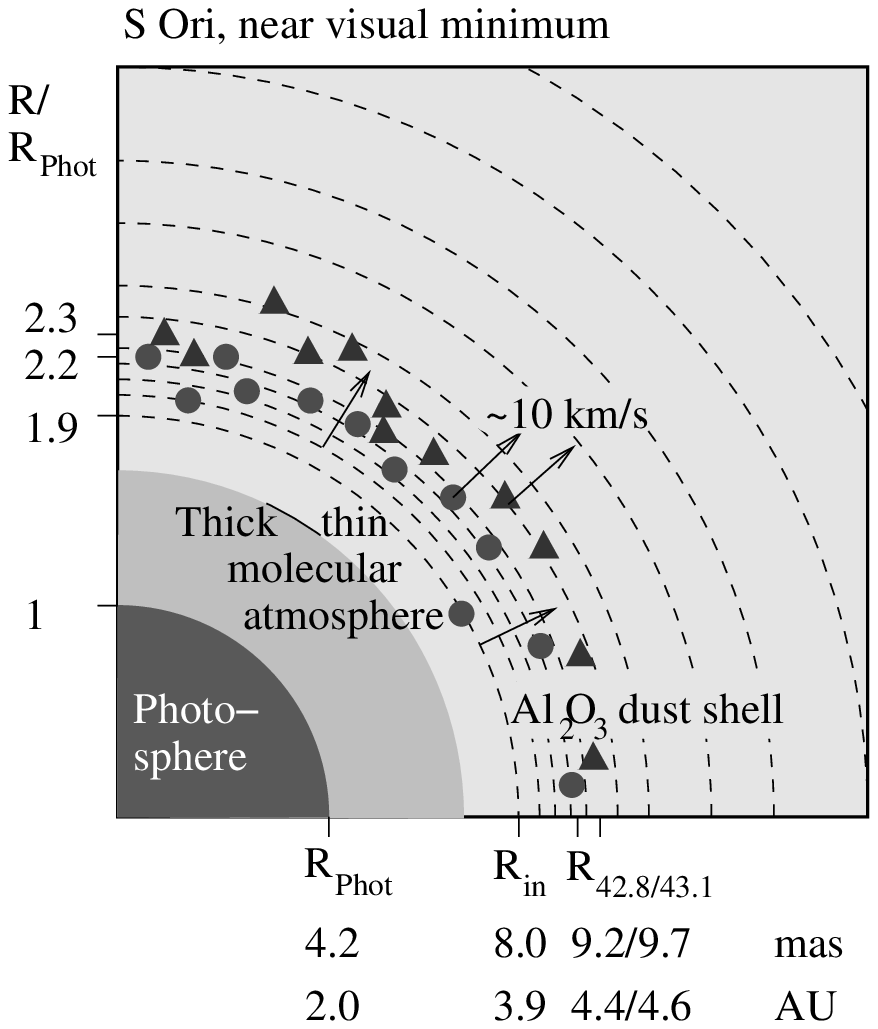}}
\resizebox{0.45\hsize}{!}{\includegraphics{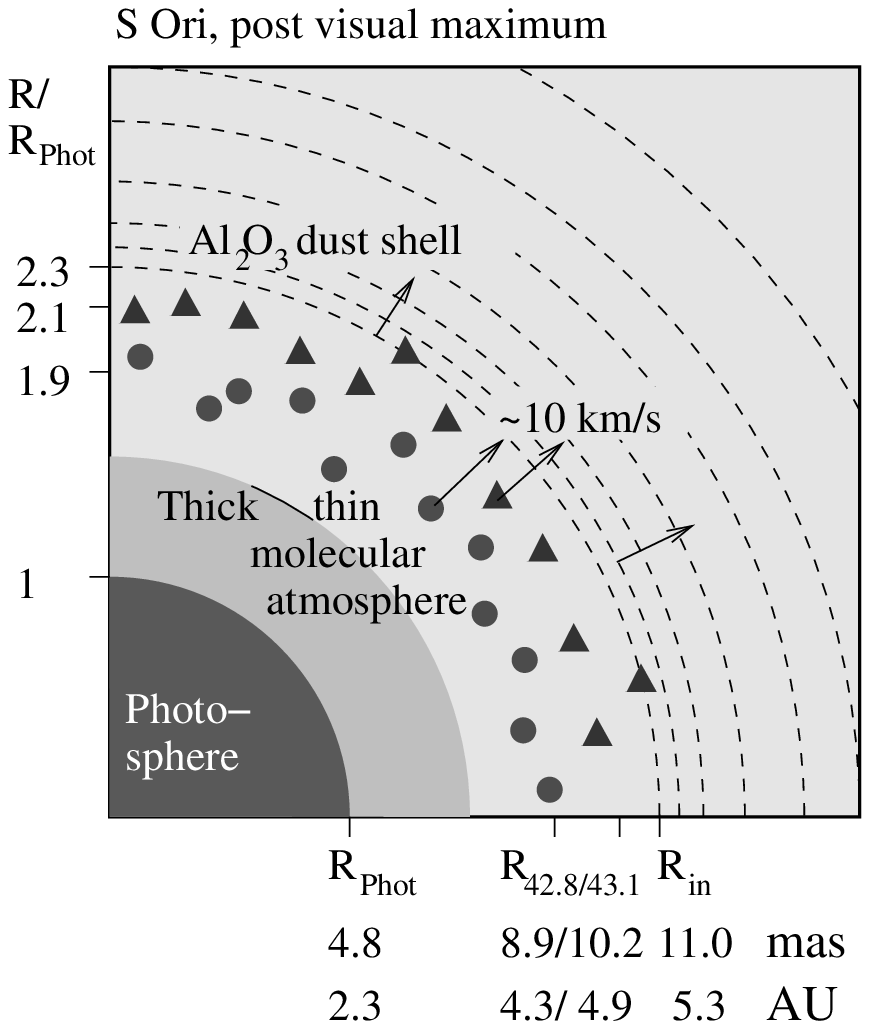}}
\caption{Sketch of the radial structure of S Ori's circumstellar
envelope at (left) near-minimum and (right) post-maximum visual phase
from Wittkowski et al. (2007). Shown are the locations of the continuum
photosphere (dark gray), the at $N$-band optically thick molecular
atmosphere (medium dark gray), the at $N$-band optically thin molecular
atmosphere (light gray), the Al$_2$O$_3$ dust shell (dashed arcs), and the
42.8 GHz/ 43.1 GHz SiO maser spots (circles/triangles).}
\end{figure}

Mid-infrared interferometric data of S Ori were taken concurrently with
additional three epochs of VLBA observations of the same SiO maser transitions
(Wittkowski et al. 2007). The modeling of the MIDI data resulted in 
phase-dependent continuum photospheric angular diameters.
The dust shell could best be modeled with Al$_2$O$_3$ grains alone
located close to the stellar photosphere with
inner radii between 1.8 and 2.4 photospheric radii.
Mean SiO maser ring radii were found to lie between about 1.9
and 2.4 stellar continuum radii.
The maser spots marked the region of the molecular
atmospheric layers shortly outward of the steepest decrease of the
mid-infrared model intensity profile. These results suggested that
the SiO maser shells are co-located with the Al$_2$O$_3$ dust shell
near minimum visual phase. Their kinematics showed that there appeared
a velocity gradient at all epochs, with masers toward the blue- and
red-shifted ends of the spectrum lying closer to the center of the 
distribution than masers at intermediate velocities. This phenomenon
was interpretated as a radial gas expansion with a 
velocity of about 10\,km/sec.
Fig.~1 shows a sketch of the radial
structure of S~Ori's circumstellar environment as derived from this study.
A similar -- but longer -- study of the Mira variable RR Aql, which
shows a silicate dust chemistry, is presented by Karovicova et al.
in these proceedings.

The combination of the dynamic model atmospheres plus dust shell model
with the maser propagation model by Gray et al. (2009) showed that 
modeled SiO masers formed
in rings with radii consistent with those found in the VLBA observations
described above and in earlier models. This agreement required the adoption
of a radio photosphere of radius about twice that of the near-infrared 
continuum photosphere in agreement with observations. Maser rings, a shock,
and the 8.1\,$\mu$m radius, dominated by optically thick water layers, appeared
to be closely related. The maser ring variability and number of spots may
not be consistent with observations, which may be explained by
re-setting masers in the model at each phase.

Near-infrared spectro-interferometric observations of AGB stars using
the AMBER instrument were first obtained by Wittkowski et al. (2008).
These observations covering 29 spectral channels between 1.29\,$\mu$m and
2.32\,$\mu$m exhibited significant variations as a function of spectral 
channel that could only be explained by a variation of the apparent angular size
with wavelength. This `bumpy' visibility curve was interpreted as a signature
of molecular layers lying above the continuum-forming
photosphere, at near-infrared wavelengths mostly CO and H$_2$O.
The variation of visibility and corresponding diameter values
resemble well the predictions by dynamic model atmospheres that naturally
include these atmospheric molecular layers. Similar bumpy visibility
curves were subsequently also seen for the red supergiant VX~Sgr 
(Chiavassa et al. 2010), the semi-regular AGB star RS Cap (Marti-Vidal et 
al., in preparation), and three OH/IR stars (Ruiz Velasco et al., these
proceedings)., indicating that close molecular layers 
may be a common phenomenon of cool evolved stars.

Medium resolution ($R$ $\sim$1500) 
visibility functions of the Mira variable R~Cnc obtained within the project
discussed here confirm the conclusion that Mira variables show 
wavelength-dependent angular diameters when observed with spectro-interferometric
techniques. In particular, the CO band-heads are nicely visible with
the AMBER medium resolution mode. The data are well consistent with predictions
by dynamic model atmospheres of the P/M as well as CODEX series, where the
latter provides a better agreement, in particular for the CO band-heads. 
R~Cnc shows 
closure phase values that are significantly different from 0\deg and 180\deg, 
thus indicate a significant deviation from point symmetry.
The interpretation of the closure phase measurements is work in progress.
They might indicate a complex non-spherical stratification
of the extended atmosphere, and may reveal whether observed asymmetries
are located near the photosphere or in the outer molecular layers. 
The measured angular diameter values together with the SAAO photometry
results in phase-dependent effective temperature values that are roughly
consistent with the effective temperature of the best-fitting model atmospheres
of the series.

\begin{figure}
\centering
\resizebox{0.9\hsize}{!}{\includegraphics{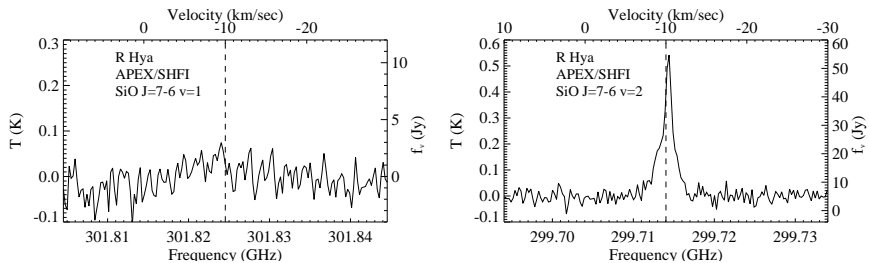}}
\caption{
APEX/SHFI observations of the $J=7-6$, $v=1$ (left)
and $v=2$ (right) SiO maser transitions of R~Hya at one epoch.
R~Hya shows only the $v=2$ transition at this phase. 
Other sources showed both transitions (e.g. R Leo), and others
showed $v=1$ but not $v=2$ (e.g. o~Cet).
}
\end{figure}

Our recent APEX observations of the $v=1$ and $v=2$ $J=7-6$ SiO maser
transition of AGB stars showed a variability of the maser intensity
that is stronger than for the centimeter SiO maser transitions. Also,
different ratios between the v=1 and v=2 transitions were detected
for different sources and phases, where even only one of the two 
transitions may be present 
The combination of dynamic atmosphere and maser
propagation models by Gray et al. (2009) showed the $v=1$ transition but
not the $v=2$ transition. Earlier such models (Humphreys et al. 2002)
showed
both transitions; these had stronger shocks and higher post-shock
temperatures compared to the more recent models. It is also known
that infrared line overlap of SiO and H$_2$O can deeply effect the
pumping of some SiO maser transitions and lead to anomalous
maser intensities (e.g. Bujarrabal et al. 1996)

\section{Summary}
We have observed a sample of AGB stars using near-infrared, mid-infrared,
and radio interferometry.
Near-infrared spectro-interferometry has shown to be a powerful tool to study
the complex atmosphere of AGB stars including atmospheric molecular layers,
most importantly H$_2$O and CO. These observations are well consistent
with predictions by recent dynamic model atmospheres.
Near-infrared closure phase measurements indicate a complex non-spherical
stratification of the atmosphere.
The addition of near-infrared photometry allows us to determine
phase-dependent effective temperature values.
Mid-infrared interferometry constrains dust shell parameters including
Al$_2$O$_3$ dust with inner boundary radii of about 2 photospheric radii
and silicate dust with inner boundary radii of about 4 photospheric radii.
SiO maser transitions observed with the VLBA (42.8\,GHz and 43.1\,GHz) 
lie in the extended atmosphere seen by near-infrared and mid-infrared
interferometry, and may be co-located with Al$_2$O$_3$ dust. Their kinematics
indicate motion such as outflow. The observed location relative to the 
stellar photosphere is consistent with predictions of combined hydrodynamic
models and maser propagation models. APEX millimeter observations indicate
high-frequency SiO masers that are located probably very close to the photosphere,
and that show strong variability. We plan to add millimeter
interferometry to this study using the Atacama Large Millimeter
Array (ALMA) in order to obtain maps of high-frequency SiO masers.


\end{document}